\documentclass[aps,prl,twocolumn,nofootinbib]{revtex4-1}

\usepackage{graphicx}
\usepackage{dcolumn}
\usepackage{bm}
\usepackage{ textcomp }
\usepackage{color}
\usepackage{amsmath}
\usepackage{amssymb}
\usepackage{xcolor}
\usepackage{hyperref}
\usepackage{easyReview}
\usepackage{mathdots}
\usepackage{float}
\usepackage{lineno}
\usepackage{array}
\usepackage{xcolor,colortbl}
\usepackage[normalem]{ulem}

\definecolor{LightBlue}{rgb}{0.8,0.8,0.8}

\usepackage{amsmath} 

\allowdisplaybreaks
\DeclareUnicodeCharacter{2212}{-}
\begin{document}
\title{Suppressed superexchange interactions in the cuprates by bond-stretching oxygen phonons}

\date{\today}

\author{Shaozhi Li}
\email[]{lishaozhiphys@gmail.com}
\affiliation{Materials Science and Technology Division, Oak Ridge National Laboratory, Oak Ridge, TN 37831, USA}
\author{Steven Johnston}
\affiliation{Department of Physics and Astronomy, The University of Tennessee, Knoxville, TN 37996, USA}
\affiliation{Institute of Advanced Materials and Manufacturing, The University of Tennessee, Knoxville, TN 37996, USA\looseness=-1} 

\begin{abstract}
We study a multi-orbital Hubbard--Su-Schrieffer-Heeger model for the one-dimensional (1D) corner-shared cuprates in the adiabatic and nonadiabatic limits using the exact diagonalization and determinant quantum Monte Carlo methods. Our results demonstrate that lattice dimerization can be achieved only over a narrow range of couplings slightly below a critical coupling $g_c$ at half-filling. Beyond this critical coupling, the sign of the effective hopping changes, and the lattice becomes unstable. We also examine the model's temperature-dependent uniform magnetic susceptibility and the dynamical magnetic susceptibility and compare them to the results of an effective spin-$1/2$ Heisenberg model. In doing so, we numerically demonstrate that the lattice fluctuation induced by the $e$-ph interaction suppresses the effective superexchange interaction. Our results elucidate the effect of bond-stretching phonons in the parent cuprate compounds in general and are particularly relevant to 1D cuprates, where strong $e$-ph interactions have recently been inferred.
\end{abstract}

\maketitle
{\it Introduction}. Numerous studies have suggested that the electron-phonon ($e$-ph) interaction plays a significant role in shaping the electronic and magnetic properties of  correlated materials like the high-temperature cuprate and iron-based superconductors~\cite{LanzaraNature2001, CukReview, JohnstonPRL2012, Lu2014, Li16219, Chaix2017, BraicovichPRR2020, LeeNature2006}, the Kitaev spin liquid candidate $\alpha\mathrm{-RuCl}_3$~\cite{HentrichPRL2018, HongPRB2021, SaiMu, Shaozhi2022}, and multiferroics~\cite{HaumontPRB2006, Dipanshu2018, PoojithaPRM2019}. For example, resonant inelastic x-ray scattering experiments on the cuprates observe harmonic lattice excitations and phonon dispersion anomalies in the vicinity of charge-density-wave excitations~\cite{Chaix2017, Li16219, BraicovichPRR2020, LinPRL2020, PengPRL2020, PengPRL2020, HuangPRX2021, TamNC2022}. These observations suggest an intimate relationship between the Cu-O bond-stretching  phonons and charge order in these materials. More recently, the authors of an angle-resolved photoemission study~\cite{ChenScience2021} on the doped cuprate spin chain Ba$_{2-x}$Sr$_x$CuO$_{3+\delta}$ have inferred the presence of a strong next-nearest-neighbor attraction, which is attributed to $e$-ph interactions \cite{WangPRL2021}.

\begin{figure}[t]
\center\includegraphics[width=0.95\columnwidth]{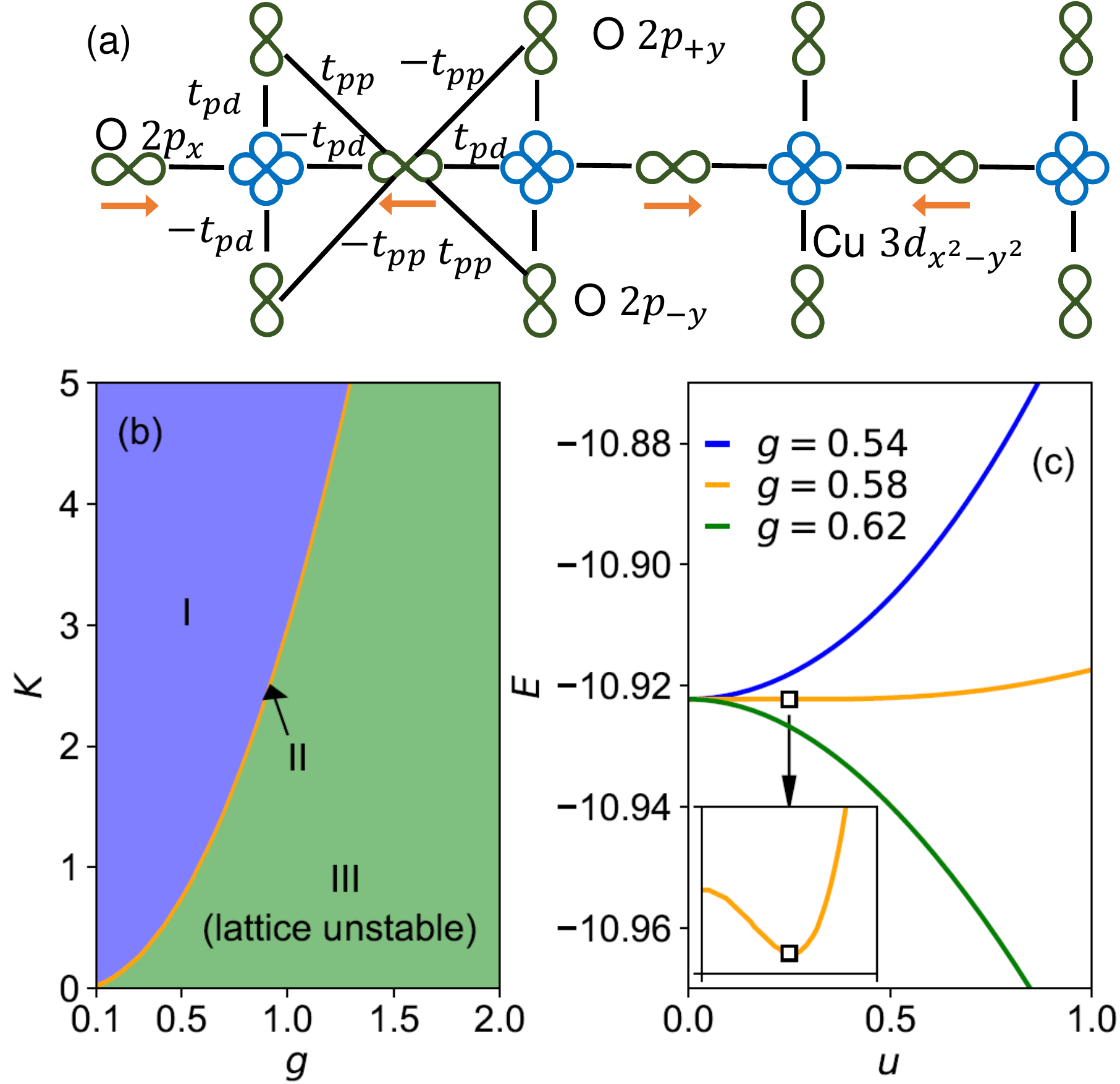}
\caption{\label{Fig:fig1} The four-orbital SSH-Hubbard model. (a) A sketch of the four-orbital Cu−O $pd$-model describing the corner-shared spin-chain cuprates like $\mathrm{Sr}_2\mathrm{CuO}_3$. The orange arrow denotes the motion of oxygen atoms for the half-breathing mode. (b) The phase diagram of the four-orbital SSH-Hubbard model in the $e$-ph coupling $g$ and elastic constant $K$ plane in the adiabatic limit. (c) The variation of energy as a function of the displacement $u$ for three different values of $g$ at $K=1$.}
\end{figure}

These observations call for further theoretical studies of Hubbard-like models with additional $e$-ph interactions. The vast majority of nonperturbative studies along these lines have focused on the Hubbard-Holstein-like model~\cite{BergerPRB1995, MatsuedaPRB2006,  KhatamiPRB2008, NowadnickPRL2012, MurakamiPRB2013, HohenadlerPRB2013, NoceraPRB2014, NowadnickPRB2015, OhgoePRL2017, MendlPRB2017, ShaozhiPRB2018, weberPRB2018, HebertPRB2019, HanPRL2020, Costa2020, CostaPRL2021}, where the lattice displacement couples to the local electron density. This coupling mechanism, however, may not relevant for the bond-stretching phonons in the cuprates, which modulate the Cu-O hopping integral and is described by a Su-Schrieffer-Heeger (SSH)-like $e$-ph interaction~\cite{SuPRL1979, Heeger1988, TangPRB1988, SenguptaPRB}. This type of interaction has been gaining interest in recent years~\cite{Hohenadler2016, SousPRL2018, BeylPRB2018, Marchand, Shaozhinpj, NoceraPRB2021, XingPRL2021, CaiPRL2021, fengarxiv}, as it may produce results quite different from the more frequently studied Holstein interaction. For example, theoretical work suggests that the SSH model may be capable of mediating high-temperature superconductivity \cite{ZhangPreprint}, enhancing antiferromagnetic (AFM) correlations in the Hubbard-SSH model~\cite{CaiPRL2021}, or stabilizing a bond-ordered phase~\cite{fengarxiv}.
However, these novel effects may be due to nonphysical changes in the sign of the effective hopping integral~\cite{ProkofevPreprint}.

Given these considerations, it is imperative to further explore the physics of Hubbard-SSH-like models. Here, we study a multi-orbital Hubbard-SSH model for quasi-1D cuprate spin chains (e.g., Ba$_{2-x}$Sr$_x$CuO$_{3+\delta}$~\cite{ChenScience2021} and Sr$_2$CuO$_3$) that retains the full Cu and O orbital degrees of freedom and the Cu-O bond-stretching motion of the lattice. We solve this model using exact diagonalization (ED) and determinant quantum Monte Carlo (DQMC), which provide numerically exact results on finite-size clusters. (Here, we can run DQMC simulations at comparatively low temperatures due to a mildler sign problem compared to 2D~\cite{ShaozhiCP2021, KungPRB2016}.) Focusing on half-filling, we find that the system has a narrow window of coupling strengths where it supports a $q = \pi/a$ bond ordering. Above this window, the lattice becomes unstable due to a sign-change in the effective hopping integral. Below this window, strong AFM correlations persist. Focusing on this physical region, we then study the magnetic properties of the model and demonstrate that the $e$-ph coupling \emph{reduces} the effective superexchange interaction.

{\it Model}. We consider a  four-orbital Hubbard-SSH model for the corner-shared cuprates that includes the Cu $3d_{x^2-y^2}$ and O $2p_{x/y}$ orbitals (the $pd$-model hereafter), as shown in Fig.~\ref{Fig:fig1}(a). The Hamiltonian is given by
\begin{eqnarray}
H&=&(\epsilon_d-\mu)\sum_{i,\sigma}\hat{n}_{i,\sigma}^d+\sum_{j,\gamma,\sigma}(\epsilon_{p,\gamma}-\mu)\hat{n}_{j,\gamma,\sigma}^{p} \nonumber\\
&+&\sum_{\substack{\langle i,j \rangle \\
 \nonumber
 \gamma,\sigma}} t_{p_\gamma d}^{ij}\left( d_{i,\sigma}^{\dagger}p_{j,\gamma,\sigma}^{\phantom\dagger} + h.c.  \right) 
+\sum_{\substack{\langle j,j^\prime \rangle\\ \gamma,\gamma^\prime,\sigma}} t_{p_\gamma p_{\gamma^\prime}}^{j j^\prime} p^{\dagger}_{j,\gamma,\sigma} p^{\phantom\dagger}_{j^\prime,\gamma^\prime,\sigma}
\nonumber \\
&+&U_d\sum_{i}\hat{n}_{i,\uparrow}^{d} \hat{n}_{i,\downarrow}^{d} 
+U_p\sum_{j,\gamma} \hat{n}_{j,\gamma,\uparrow}^{p}\hat{n}_{j,\gamma,\downarrow}^{p} \nonumber\\
&+&\sum_{j} \left(\frac{\hat{P}_j^2}{2M} + K \hat{x}_j^2  \right)  .\label{Eq:Hpd}
\end{eqnarray}
Here, $\langle \cdots \rangle$ represents a sum over nearest neighbor orbitals; $d^{\dagger}_{i,\sigma}$  and $p^\dagger_{j,\gamma,\sigma}$ creates a hole with spin $\sigma$ ($=\uparrow,~\downarrow$) on the $i^{\mathrm{th}}$ Cu $3d_{x^2-y^2}$ and the $j^\mathrm{th}$ O $2p_\gamma$ ($\gamma=x,\pm y$) orbitals, respectively; $\epsilon_d$ and  $\epsilon_{p,\gamma}$ are the onsite energies; $\hat{n}_{i,\sigma}^{d}$ and  $\hat{n}_{j,\gamma,\sigma}^p$ are the number operators for the Cu $3d_{x^2-y^2}$ orbital and O $2p_\gamma$ orbital, respectively;  $\hat{x}_j$ and $\hat{P}_j$ are the displacement and momentum operators for the $2p_x$ orbital; $g$ is the $e$-ph coupling strength.
$t_{p_\gamma  d}^{ij}$ and $t_{p_\gamma p_{\gamma^\prime}}^{j j^\prime}$ are the nearest-neighbor Cu-O ($2p_{\gamma}$) and O-O hopping integrals.
We assume that the vibration of the oxygen atom linearly modulates the magnitude of the hopping $t_{p_xd}^{ij}(x_j)=t_{p_xd}^{ij}(1-Q_{\pm}g\hat{x}_j)$, where $Q_\pm = \pm1$ for hopping to the left/right. Note, however, that all hopping integrals retain the phase convention shown in Fig.~\ref{Fig:fig1}(a); $U_d$ and $U_p$ are the onsite Hubbard interactions on the Cu and O orbitals, respectively; $\mu$ is the chemical potential, which is used to adjust the hole density. $M$ and $K$ are the mass and elastic constant, and the phonon frequency is given by $\omega_{\mathrm{ph}}=\sqrt{\frac{2K}{M}}$. 
Throughout this work, we study the model at half-filling (1 hole/Cu) and set (in units of eV) $\epsilon_d=0$, $\epsilon_{p,x}=3$, $\epsilon_{p,y}=3.5$, $|t_{p_xd}|=1.5$, $|t_{p_yd}|=1.8$, $|t_{pp}|=0.75$, $U_d=8$, $U_p=4$. The periodic boundary condition is used in this work.

\begin{figure}[t]
\center\includegraphics[width=\columnwidth]{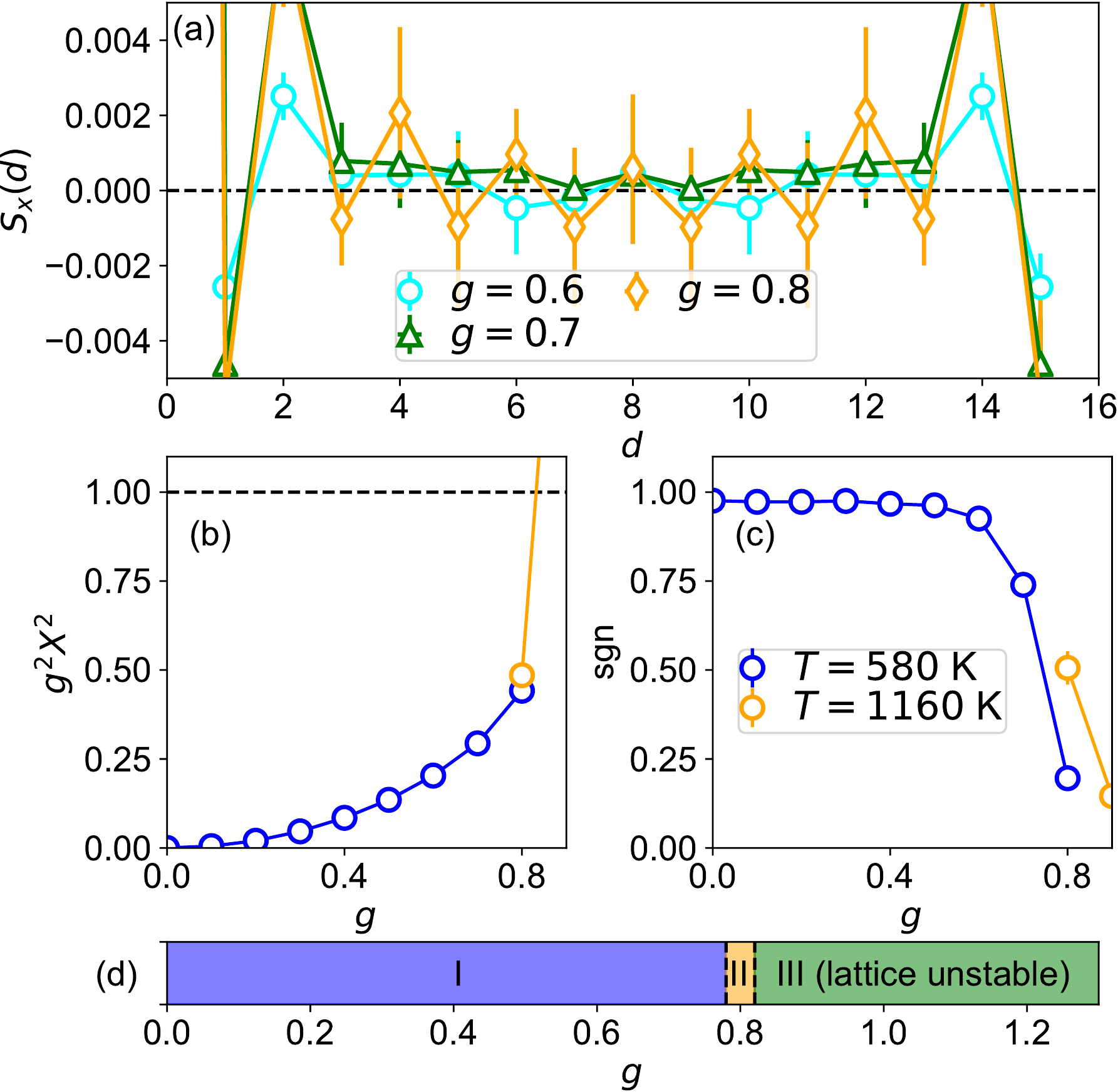}
\caption{\label{Fig:fig2} (a) The displacement correlation function $\langle S_x(d) \rangle$ for three different values of $g$ at $T=580$ K. (b) The variation of $g^2X^2$ as a function of the $e$-ph coupling strength $g$. (c) The sign value as a function of $g$. (d) A sketch of the pseudo phase diagram as a function of $g$.}
\end{figure}

\begin{figure}[t]
\center\includegraphics[width=\columnwidth]{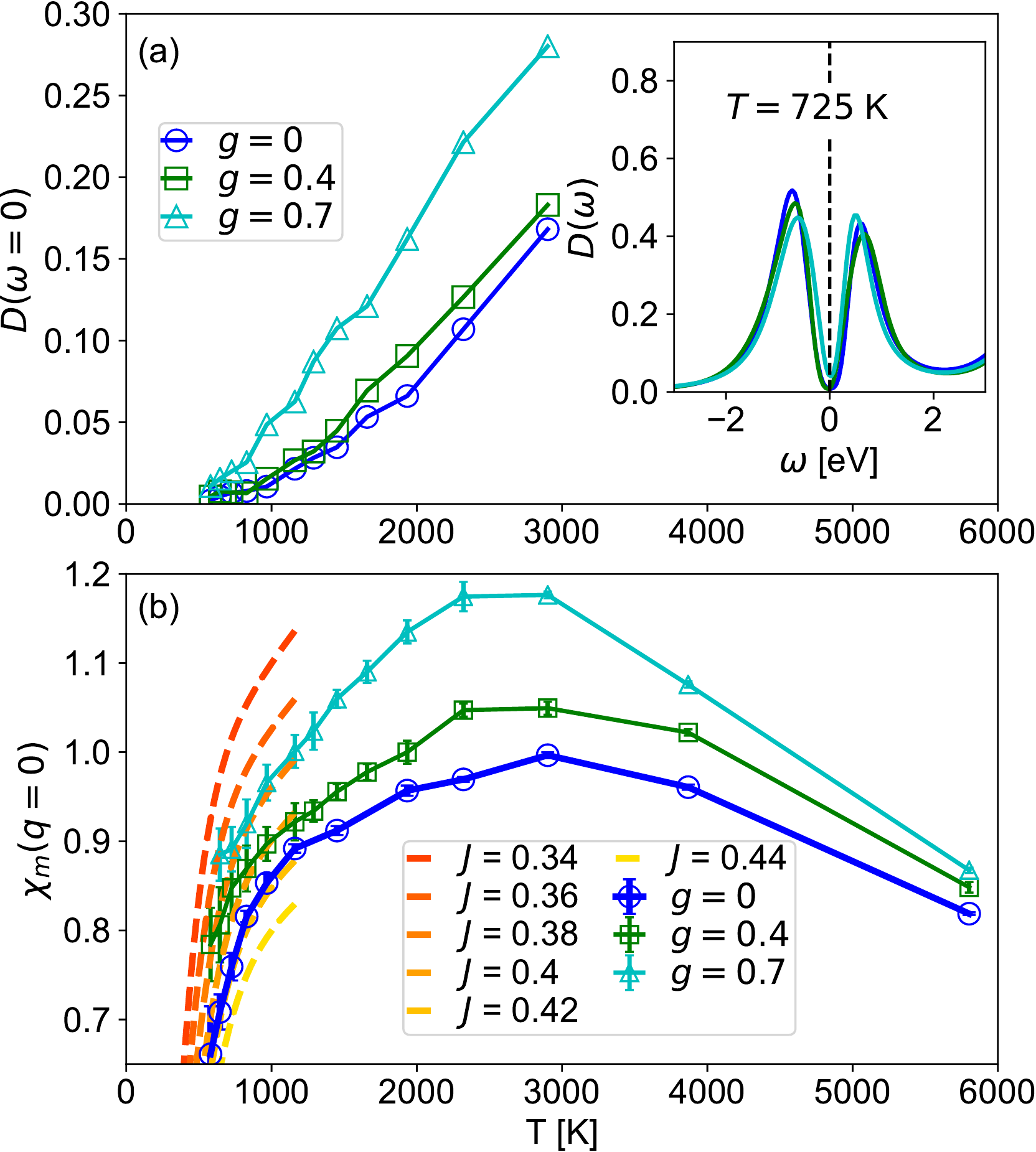}
\caption{\label{Fig:fig3} (a) The density of states $D(\omega)$ as a function of temperature for three different $e$-ph coupling strengths $g$. The inset plots the density of states in the frequency space at $T=725$ K. (b) The uniform magnetic susceptibility $\chi_m(q=0)$ as a function of temperature. The dashed curves show the results of the Heisenberg model. }
\end{figure}

Our model includes the motion of 2$p_x$ orbitals, which is more essential to describe the electronic structure along the chain direction. For a two-dimensional system, the motion of the 2$p_y$ orbital should also be considered.

{\it Adiabatic limit}. We first examine the adiabatic limit $\omega_\mathrm{ph}\rightarrow 0$, where phonons can be treated as classical fields. In this case, we consider the half-breathing mode and parameterize $x_j=(-1)^ju$. We then obtain the ground state by minimizing the system's energy $E(u)$ with respect to $u$, which is done by performing ED on a Cu$_4$O$_{12}$ cluster. Figure~\ref{Fig:fig1}(b) plots the resulting $K$-$g$ phase diagram, where we find three distinct solutions each characterized by their $E(u)$ behavior [see Fig.~\ref{Fig:fig1}(c)]. In region I, the chain has strong antiferromagnetic correlations, and $E(u)$ is a monotonically increasing function of $u$ such that the ground state has no half-breathing distortion ($u = 0$). In region II, $E(u)$ develops a local minimum at small $u$, and the ground state has a small half-breathing distortion leading to a dimerized state. We note that the energy well for this state is very shallow, implying that the dimerization fluctuates significantly even at low temperatures. In region III, appearing at large $e$-ph coupling strength, the total energy $E(u)$ monotonically decreases as $u$ increases, indicating that the lattice is unstable. The boundary for this region is determined by the condition $|gu|>1$, which corresponds to the boundary where the effective hopping integral changes sign.  

{\it Nonadiabatic case}. Next, we study a nonadiabatic case at finite temperatures with $\omega_\mathrm{ph}=\sqrt{2}$ eV and $L=16$ unit cells using DQMC. (For details of the method, we refer the reader to Refs.~\cite{ShaozhiCP2021} and \cite{Shaozhinpj}.) We now include the individual atomic motion of each 2$p_x$ orbital along the chain direction. 

Figure~\ref{Fig:fig2}(a) plots the displacement correlation function $S_x(d) = \frac{1}{L}\sum_{i}\langle \hat{x}_{i+d}\hat{x}_{i} \rangle$ at $T=580$ K. At $g=0.6$ and $g=0.7$, the $q = \pi$ correlations are short-ranged; however, for $g= 0.8$, they grow stronger and extend across the entire cluster. We note, however, that the error bars are larger than the mean value, indicating that the lattice displacements are fluctuating significantly at this temperature. Figure~\ref{Fig:fig2}(b) shows the variation of $g^2x^2$ as a function of $g$. This quantity increases with $g$ when $g\le 0.8$ but rapidly jumps above one when $g>0.8$, implying that the lattice is unstable for these larger couplings. Note, the results for $g>0.8$ are obtained at $T=1160$ K because the average fermion sign for $g=0.9$ and $T=580$ K is extremely small [see Fig.~\ref{Fig:fig2}(c)]. Based on these results, it is difficult to determine whether the strong Peierls fluctuation at $g = 0.8$ will give way to a long-range ordered state at a lower temperature. Even though the dimerized state is not observed in our results, we can qualitatively divide the parameter space into three regions, shown in Fig.~~\ref{Fig:fig2}(d). In region I ($g<0.78$), the system has strong antiferromagnetic fluctuations; In region II ($0.78<g<0.82$), the system has antiferromagnetic and dimerization fluctuations; Finally, in region III ($g>0.82$), the system is unstable. 

{\it Finite-temperature magnetic correlations}. The $e$-ph interaction in the Hubbard-Holstein model drives charge-density-wave correlations that compete directly with antiferromagnetism~\cite{BergerPRB1995, MatsuedaPRB2006,  KhatamiPRB2008, NowadnickPRL2012, MurakamiPRB2013, HohenadlerPRB2013, NoceraPRB2014, NowadnickPRB2015, OhgoePRL2017, MendlPRB2017, ShaozhiPRB2018, weberPRB2018, HebertPRB2019, HanPRL2020, Costa2020, CostaPRL2021}. A recent study of the singleband Hubbard-SSH model in 2D~\cite{CaiPRL2021}, however, has concluded that antiadiabatic phonons can enhance antiferromagnetism in this model. We, therefore, also examined how the SSH interaction renormalizes the electronic and magnetic properties of our model. 

Figure~\ref{Fig:fig3}(a) plots the density of states (DOS) at the Fermi level $D(\omega=0)$ as a function of temperature. 
For $g=0$, $D(0)$ approaches zero around $T\approx967$~K, indicative of a metal-to-insulator transition. The inset of Fig.~\ref{Fig:fig3}(a) shows the DOS over a wider energy interval at $T = 725$ K, which also exhibits a clear gap at $\omega = 0$. When $g\ne 0$, the metal-to-insulator transition shifts to lower temperatures, and the corresponding Mott gap begins to narrow.

To gain insight into the effect of bond-stretching phonons on the magnetic properties, we calculate the uniform magnetic susceptibility $\chi_m(q = 0)=\frac{1}{T}\sum_{i}\langle \hat{S}^{z}_{i}\hat{S}^{z}_{0}\rangle$. Here, $\hat{S}_{i}^{z}=\sum_{\gamma} \hat{S}^z_{i,\gamma}$ is the total spin operator for the unit cell~\cite{ShaozhiCP2021}. For a spin $1/2$ system like $\mathrm{Sr}_2\mathrm{CuO}_3$, the uniform magnetic susceptibility is frequently used by experimentalists to determine the value of the effective superexchange coupling $J$~\cite{BonnerPR1964, GriffithsPR1964, EggertPRL1994, MotoyamaPRL1996}. We apply the same methodology here. Figure~\ref{Fig:fig3}(b) plots $\chi_m$ for our model as a function of temperature for several values of $g$. For comparison, the dashed curves represent results obtained from the Heisenberg model defined on an $L = 16$ site chain with different values of $J$. For $g=0$, the low-temperature behavior of $\chi_m$ overlaps with Heisenberg model predictions when $J=0.42$ eV. This observation implies that a simple Heisenberg model can capture the magnetic behavior of the multiorbital system for our choice of parameters. Increasing the $e$-ph interaction increases $\chi_m$ of the $pd$-model, implying that the effective superexchange interaction is suppressed. We also find that the $pd$-model's $\chi_m$ for $g=0.4$ and $g=0.7$ deviate from predictions of the Heisenberg model at low temperatures. This discrepancy likely originates from the increased role of the oxygen atoms at these coupling strengths or the spin-phonon interaction.

We can understand the suppression of the superexchange interaction using a mean-field analysis. At half-filling, $J$ can be obtained from the perturbation theory and is given by $J=\frac{t_{pd,1}^2 t_{pd,2}^2}{\Delta_{dp}^2}(\frac{1}{\Delta_{dp}+U_{p}/2}+\frac{1}{U_d})$~\cite{Hanke2010}. Here, $\Delta_{dp}=\epsilon_d-\epsilon_{p_x}$, and $t_{pd,1/2}$ is the hopping integral between $2p_x$ and the left/right $3d_{x^2-y^2}$  orbitals. When the oxygen atom moves away from its equilibrium position, $J$ becomes $J(1-g^2x^2)^2$. Consequently, \emph{any} bond-stretching of oxygen atoms suppresses the effective superexchange interaction. While this analysis neglects potential phonon-induced renormalization of the Hubbard interaction, it does demonstrate that the modulation of hopping integrals induced by the lattice displacement suppresses the effective superexchange interaction.

\begin{figure}[t]
\center\includegraphics[width=\columnwidth]{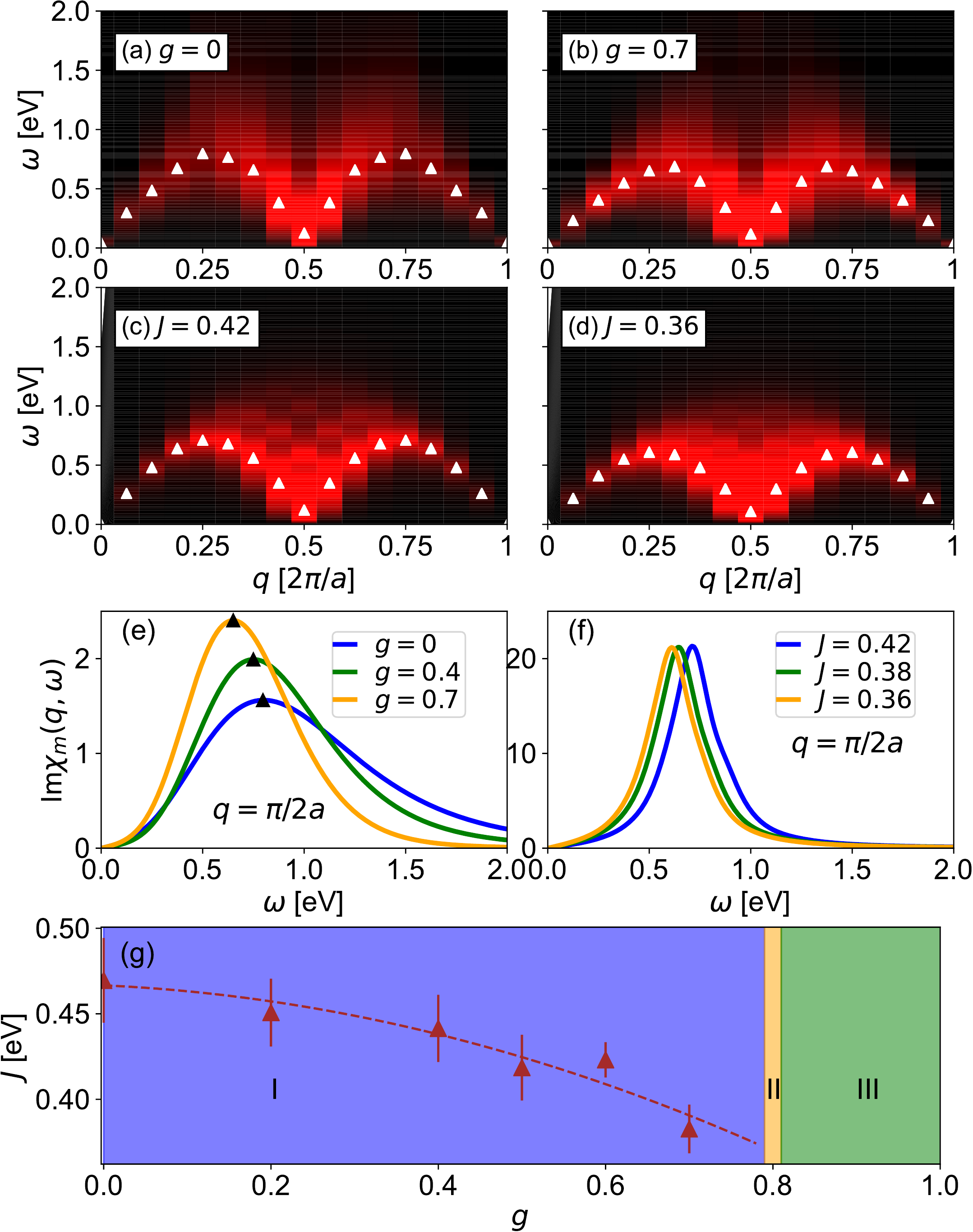}
\caption{\label{Fig:fig4} The imaginary part of the dynamical magnetic susceptibility $\mathrm{Im}\chi_m(q,\omega)$ at $T=645$ K. Panels (a) and (b) show DQMC results for $\mathrm{Im}\chi_m(q,\omega)$ of the $pd$-model with an $e$-ph coupling strength $g=0$ and 0.7, respectively. Panels (c) and (d) show ED results $\mathrm{Im}\chi_m(q,\omega)$ of the Heisenberg model for $J=0.42$ and 0.36, respectively. Panels (e) and (f) plot  $\mathrm{Im}\chi_m(q,\omega)$ at $q=\frac{\pi}{2a}$ for the $pd$- and Heisenberg models, respectively. Panel (g) plots the estimated effective superexchange interaction of the $pd$-model as a function of the $e$-ph coupling strength.
}
\end{figure}

Next, we analyze the imaginary part of the dynamical magnetic susceptibility $\mathrm{Im}\chi_m(q,\omega)$, which is plotted in Fig.~\ref{Fig:fig4}. Here, panels (a) and (b) plot DQMC results of the $pd$-model, while panels (c) and (d) show ED results of the 1D Heisenberg model. 
The white triangles denote the frequency $\omega_m(q)$ of the maximum of the spectra at each momentum point. Our system is a charge-transfer insulator with robust antiferromagnetic correlations at low temperature and half-filling. Its magnetic excitation is therefore characterized by a  two-spinon continuum \cite{KarbachPRB1997, Jean2006}, which is observed in both the $pd$- and Heisenberg model results. Compared to the results at $g=0$, $\omega_m(q)$ of the $pd$-model is smaller at $g=0.7$. For better visualization, Fig.~\ref{Fig:fig4}(e) plots $\mathrm{Im}\chi_m(q,\omega)$ for $g=0$, $0.4$, and $0.7$ at $q=\frac{\pi}{2a}$. Here, we observe that the peak in the susceptibility softens to lower energy as $g$ increases, reflecting an overall reduction in the effective superexchange interaction. Figure~\ref{Fig:fig4}(f) shows that the effective Heisenberg model captures the softening but does not capture the redistribution of the spectral weight. This inconsistency occurs because the effective Heisenberg model cannot correctly describe the distribution of the local moment onto the oxygen sites, as discussed in our previous work~\cite{ShaozhiCP2021}.

As an alternative to fitting the uniform magnetic susceptibility, we can estimate the effective $J$ by adjusting the interaction in the Heisenberg model to fit $\omega_m(q)$ of the $pd$-model. Figure~\ref{Fig:fig4}(g) plots the estimated superexchange interaction $J$ extracted from our DQMC simulations as a function of $g$. At $g=0$, the superexchange interaction is about 0.46 eV, which is slightly larger than the value shown in Fig.~\ref{Fig:fig3}(b) ($J=0.42$ eV). Due to the thermal broadening and the Maximum Entropy method, the spectra have a broad peak. We, therefore, provide approximate error bars for our fits, which are estimated as the energy range over which $\mathrm{Im}\chi_m(q,\omega)\ge 0.99 \mathrm{Im}\chi_m(q,\omega_m(q))$. As the $e$-ph coupling strength increases in region I of our phase diagram, the effective superexchange interaction decreases.

{\it Summary}. We studied the effects of bond-stretching phonons in a four-orbital Hubbard-SSH model for cuprate spin chains. In the adiabatic limit and at zero temperature, the model only supports a dimerized state in an extremely narrow window of $e$-ph coupling strengths. In the nonadiabatic case, the dimerized state is replaced by strong lattice fluctuations at a finite temperature. In both cases, we found that the lattice becomes unstable when the $e$-ph coupling exceeds a critical value that roughly corresponds to the point where the effective hopping changes sign. Our results suggest that achieving a dimerized state in the parent cuprate compounds is difficult based on the analysis of the SSH model. This difficulty arises from the limitation of the linear SSH model, which does not capture the giant repulsion between atoms that occurs when they are too close.  

We also examined the $pd$-model's electronic and magnetic properties as a function of $e$-ph coupling strengths. By mapping the $pd$-model's magnetic properties to an effective Heisenberg model, we find that the effective superexchange interaction decreases quadratically with the $e$-ph coupling strength in the region where the linear SSH interaction is physical. Our results, therefore, speak against the idea that these interactions can enhance AFM correlations in cuprate materials \cite{CaiPRL2021}. For the Hubbard-Holstein model, the $e$-ph interaction decreases the effective Coulomb interaction. In the large $U$ limit, where $J\approx -\frac{4t^2}{U}$, a weak $e$-ph coupling can increase the effective superexchange interaction. Therefore, the $e$-ph interaction plays different roles in modifying the magnetic properties in the Hubbard-Holstein and four-orbital Hubbard-SSH models. These considerations highlight the need to retain the oxygen orbitals when considering SSH-like interactions in materials like the cuprates.

\section{Acknowledgements}
The authors thank Mona Berciu, Benjamin Cohen-Stead, and Richard Scalettar for useful discussions and comments on this work. 
S.L. was supported by the U.S. Department of Energy, Office of Basic Energy Sciences, Materials Sciences and Engineering Division. S.J. was supported by the U.S. Department of Energy, Office of Science, Office of Basic Energy Sciences, under Award Number DE-SC0022311. This research used resources of the Compute and Data Environment for Science (CADES) at the Oak Ridge National Laboratory, which is supported by the Office of Science of the U.S. Department of Energy under Contract No. DE-AC05-00OR22725. 

\bibliography{main.bbl}

\end{document}